\documentclass[prx,floatfix,showpacs,superscriptaddress,longbibliography,twocolumn]{revtex4-1}

\usepackage[USenglish]{babel}

\usepackage{amsmath,amssymb,amsfonts}

\usepackage{bm}

\usepackage{epsfig}
\usepackage{subfigure}

\usepackage[colorlinks=true,linkcolor=blue,citecolor=red]{hyperref}

\newcommand{\equ}[1]
{Eq.~(\ref{#1})}

\newcommand{\figu}[1]
{Fig.~\ref{#1}}

\newcommand{\appu}[1]
{App.~\ref{#1}}

\def\=={\equiv}

\def\cG0{{\cal G}_0} 
\def\cG{{\cal G}}

\def\=={\equiv}

\newcommand{\be}{\begin{equation}}
\newcommand{\ee}{\end{equation}}

\begin{document}

\date{\today}

\author{Francesco~Grandi}
\affiliation{Department of Physics, University of Erlangen-N\"urnberg,
  91058 Erlangen, Germany}

\author{Jiajun~Li}
\affiliation{Department of Physics, University of Erlangen-N\"urnberg,
  91058 Erlangen, Germany}

\author{Martin~Eckstein}
\affiliation{Department of Physics, University of Erlangen-N\"urnberg,
  91058 Erlangen, Germany}

\title{Ultrafast Mott transition driven by nonlinear electron-phonon interaction}

\begin{abstract}

Nonlinear phononics holds the promise for controlling properties of quantum materials on the ultrashort timescale. 
Using nonequilibrium dynamical mean-field theory, we solve a model for the description of organic solids, where correlated electrons couple non-linearly to a quantum phonon-mode. 
Unlike previous works, we exactly diagonalize the local phonon mode within the non-crossing approximation (NCA) to include the full phononic fluctuations. 
By exciting the local phonon in a broad range of frequencies near resonance with an ultrashort pulse, we show it is possible to induce a Mott insulator-to-metal phase transition. 
Conventional semiclassical and mean-field calculations, where the electron-phonon interaction decouples, underestimate the onset of the quasiparticle peak. 
This fact, together with the non-thermal character of the photo-induced metal, suggests a leading role of the phononic fluctuations and of the dynamic nature of the state in the vibrationally induced quasiparticle coherence.
\end{abstract}
\maketitle


\textit{Introduction -} 
In the last decade, nonlinear phononics \cite{Forst2011,PhysRevB.89.220301} has become 
one of the most promising pathways for the non-equilibrium control of quantum materials \cite{Basov2017}. 
Within this approach, one can transiently stabilize a crystal structure unstable under 
equilibrium conditions by coherently exciting an infrared-active lattice mode which is nonlinearly 
coupled to a Raman-active phonon \cite{Rini2007N,PhysRevLett.108.136801,Mankowsky2014,Forst2011PRB,Forst2015,Nova2017,PhysRevB.95.024304}. 
An analogous pathway suggests that the electronic properties of solids may be manipulated by the excitation of 
vibrational modes that couple non-linearly with the local degrees of freedom of the electronic system 
\cite{PhysRevLett.115.187401,Kennes2017_NatPhys,PhysRevB.95.205111,2020arXiv200313447M,PhysRevB.98.165138,PhysRevResearch.2.013336}.

Molecular solids provide a perfect playground for testing this mechanism \cite{Lang2004,doi:10.1143/JPSJ.81.011004}. 
A paradigm example is the charge transfer salt ET-F$_{2}$TCNQ, which is an 
archetypical one-dimensional Mott insulator under equilibrium conditions and has been widely studied under photo-doping \cite{PhysRevLett.98.037401,PhysRevLett.112.117801}. 
Upon excitation of the molecular vibration $\omega_{\text{ph}} = 1000 \ \mathtt{cm}^{-1}$, 
the charge transfer (CT) resonance at $\sim 5500 \ \mathtt{cm}^{-1}$ is red-shifted, 
and an in-gap state at $2\times\hbar \omega_{\text{ph}}$ appears \cite{Kaiser2014_SciRep,PhysRevLett.115.187401}. 
Even more intriguing results have been obtained regarding light-induced 
superconductivity. The fulleride K$_{3}$C$_{60}$ \cite{RevModPhys.81.943} shows a 
superconducting state at temperatures almost ten-times higher than the equilibrium $T_{c}$ 
lasting few-picoseconds after excitation in the frequency range related to the 
T$_{1\text{u}}$ mode of the C$_{60}$ molecule \cite{Mitrano2016,Cantaluppi2018,2020arXiv200212835B_budden}. 
The organic superconductor $\kappa-\left(\text{BEDT}-\text{TTF}\right)_{2} \text{Cu} \left[ \text{N} \left( \text{CN} \right)_{2} \right] \text{Br}$ 
(henceforth $\kappa-\text{Br}$) displays a similar behavior when the C$=$C stretching 
mode of the $\left( \text{BEDT}-\text{TTF} \right)^{+0.5}$ molecule is excited \cite{buzzi2020photomolecular}. 
Because the photo-induced superconducting response is correlated with the 
presence of an equilibrium coherent quasiparticle,
it is an obvious relevant question, which will be analyzed in this paper, whether non-linear 
phononics can enhance quasiparticle coherence in correlated electron systems. 
Moreover, we might ask if it is possible at all to replace the non-linear phonons with some time-dependent electronic Hamiltonian parameter. The rationale beyond this idea comes from a ``natural'' decoupling of the electron-phonon interaction. A typical local electron-phonon interaction 
$g O_i X_i$, where an electronic operator $ O_{i}$ couples to the displacement 
$ X_i$ of atom $i$, would give a term $F_i(t) O_i$ with a time-dependent ``force'' 
$F_i (t)=g \langle X_i (t)\rangle$ in the electronic Hamiltonian when $ X_i$ is 
replaced by its time-dependent expectation value in a coherent (macroscopically 
occupied) $\bm q=0$ phonon state. For example, the experiments on ET-F$_{2}$TCNQ 
and $\kappa-\text{Br}$ were interpreted by an average shift and a periodic 
time-dependence of the local Coulomb interaction (the Hubbard $U$) \cite{PhysRevLett.103.066403,buzzi2020photomolecular,PhysRevLett.115.187401}. 
However, it is clear that the approximation $ O_i X_i \approx O_i \langle X_i (t) \rangle$ 
is controlled by the fluctuations of the local displacement operator $ X_i$ relative to 
$\langle X_i (t) \rangle$, which are significant even when the phonons are globally in 
a macroscopic coherent mode. Even though the decoupling approach 
can successfully capture some experimental observations, 
it is, therefore, important to understand how quantum effects become manifest in 
nonlinear electron-phononics \cite{Kaiser2014_SciRep}.

In this paper, we study a generalization of the Hubbard-Holstein model \cite{PINCUS197251,PhysRevLett.87.206402}, 
which applies to ET-F$_{2}$TCNQ and $\kappa-\text{Br}$, with a quadratic 
coupling of a local displacement $X_{i}$ to the doublon and the holon densities, 
so that, on average, $X_{i}^2$ modulates a local electron interaction $U$. The 
simulations indeed predict an insulator-to-metal transition (IMT), with a strong enhancement 
of the quasiparticle weight driven by the excitation of the local vibration. The 
vibrationally induced metallicity in the model cannot be accounted for by a 
decoupled Hamiltonian in which a time-dependent $U$ replaces the effect of 
the phonons, but instead the full quantum dynamics must be taken into account. 
We show that, even if the decoupled model takes into account the back-action of the electrons on the phonons (and vice versa), it cannot capture phonon-related features in the electronic spectrum (Fano resonances) \cite{PhysRevLett.110.106401,PhysRevLett.85.5304}.


\begin{figure}
\centering \includegraphics[width=0.5\textwidth]{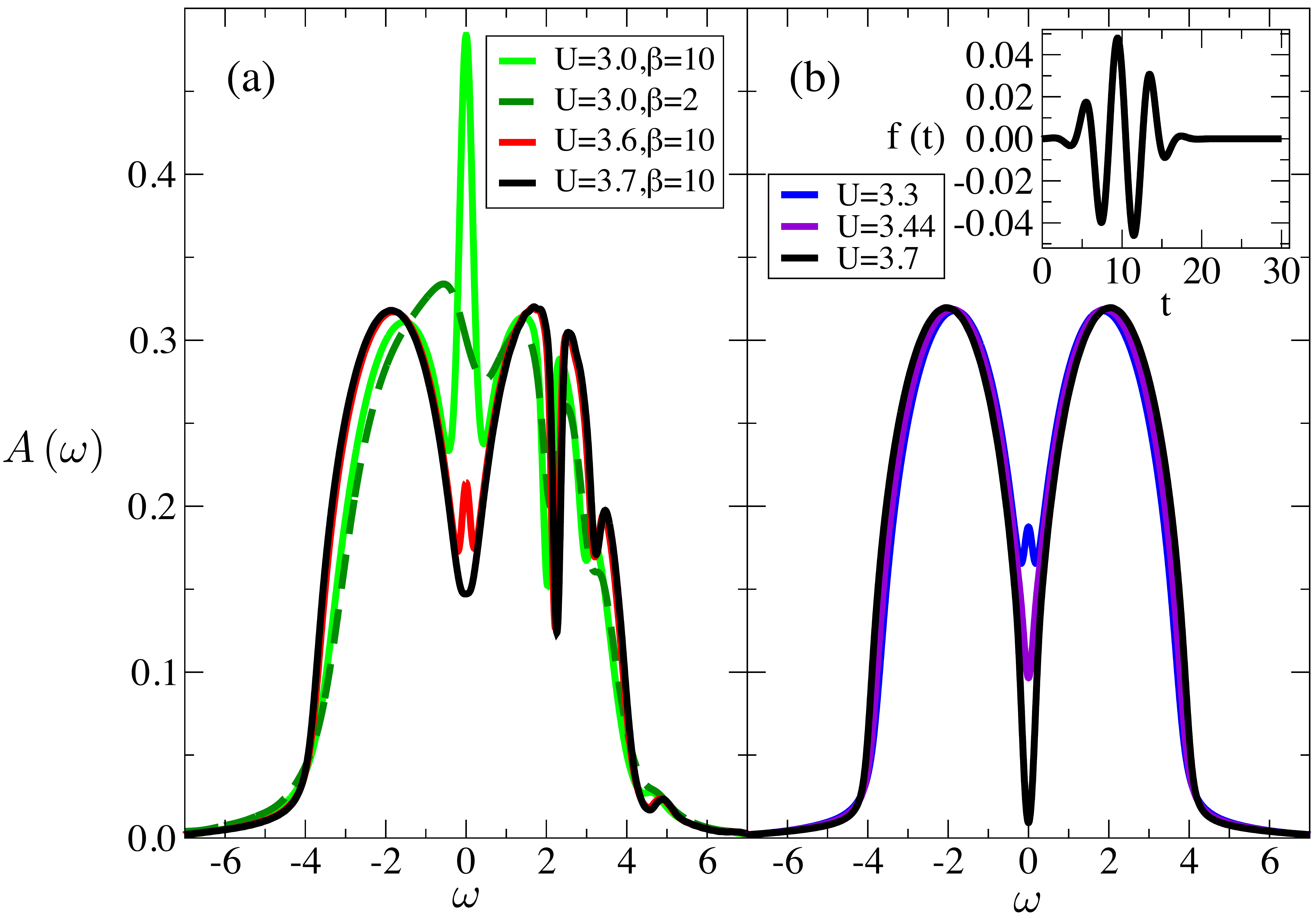}
\caption{(Color online) (a) Equilibrium spectral functions $A \left( \omega \right)$ for the electron-phonon coupled system at several values of $U$ and inverse temperature $\beta$. (b) Same for the Hubbard model without dynamic phonons, at $\beta=10$. Inset: time dependent driving field $f \left( t \right)$, cf. \equ{ham_driv}. We keep the maximum amplitude ($0.05$) and the duration ($20$), while the frequency $\Omega$ is changed ($\Omega = 1.5$ in the plot, equal to $\omega_{\text{ph}}$).}
\label{fig_1}
\end{figure}

\textit{Model -} We consider the generalized Hubbard-Holstein model, compactly 
written as:
\be \label{ham}
H \left( t \right) = H_{\text{el}} + H_{\text{el-ph}} + H_{\text{ph}} + H_{\text{driv}} \left( t \right) 
-\mu N\;.
\ee
The purely electronic part of the Hamiltonian is given by
\be 
\label{ham_el}
H_{\text{el}} = -v_{0} \sum_{\langle i,j \rangle, \sigma} \left( c^{\dagger}_{i, \sigma} c_{j, \sigma} + \text{H. c.} \right) 
 + U \sum_{i} n_{i, \uparrow} n_{i, \downarrow},
\ee
with nearest-neighbor hopping $v_0$ and local Hubbard interaction $U$; $\mu$ 
is the chemical potential used to fix the occupation of each site to $\langle n_{i} \rangle = 1$. 
$v_{0} = 1$ and $\hbar / v_{0}$ set the units for the energy and the times, respectively. 
The bare phonon Hamiltonian in \equ{ham} describes a band of Einstein phonons 
$H_{\text{ph}} = \omega_{\text{ph}} \sum_{i} (a_{i}^\dagger a_i + \frac{1}{2})$ where 
$a^{\dagger}_{i}$ ($a_{i}$) are the creation (destruction) operator for a boson.

The electron-phonon interaction is given by \cite{Kaiser2014_SciRep}:
\be \label{ham_el_ph}
H_{\text{el-ph}} = 2 \sum_{i} \left( h H_{i} - d D_{i} \right) X_{i}^{2} \;,
\ee
where $H_{i} = \left( 1 - n_{i,\uparrow} \right) \left( 1 - n_{i,\downarrow} \right)$ ($D_{i} = n_{i,\uparrow} n_{i, \downarrow}$) 
are the holon (doublon) operators, and $X_{i} = \frac{a_{i}^{\dagger}+a_{i}}{\sqrt{2}}$ is 
the local displacement.

We take parameters $\omega_{\text{ph}} = 1.5$, 
so that we are far from the adiabatic regime, and assume the 
values $h=0.1$ and $d=0.35$ for the coupling constants. 
With this, the renormalized phonon frequencies on an isolated 
site occupied by a holon or doublon are $\omega_{\text{ph}}^{h} \sim 1.24 \ \omega_{\text{ph}}$ 
and $\omega_{\text{ph}}^{d} \sim 0.26 \ \omega_{\text{ph}}$, 
respectively. The renormalized frequency values obtained this 
way are in qualitative agreement with what found for ET-F$_{2}$TCNQ, 
where a stiffening of the holon oscillator and a slackening of 
the doublon one is observed \cite{Kaiser2014_SciRep}. 
Note that $H_{\text{el-ph}} $ 
with $d\neq h$ breaks the particle-hole symmetry even on a bipartite lattice \cite{doi:10.1142/S021797920302079X}. 
In experiments, the quadratic coupling of the phonon-displacement and the electrons is 
confirmed by the $2 \omega_{\text{ph}}$ feature in the optical absorption after phonon 
excitation \cite{Kaiser2014_SciRep}. The last term in \equ{ham} describes a linear 
coupling of the phonon displacement to an external electric field,
\be \label{ham_driv}
H_{\text{driv}} \left( t \right) = \sqrt{2} \omega_{\text{ph}} f \left( t \right) \sum_{i} X_{i} \;,
\ee
and we use a few-cycle excitation pulse $f(t)$ with different frequencies $\Omega$ 
(see \figu{fig_1} inset) to excite a coherent vibration of the phonon. Finally, we will compare 
the dynamics obtained from \equ{ham} with a decoupled Hamiltonian, in which the 
electron dynamics is determined by \equ{ham_el} with a time-dependent interaction $U$
\be \label{ham_u_driv}
H_{\text{U-driv}} \left( t \right) = H_{\text{el}} [U\to U(t)] \;,
\ee
with different forms of $U(t)$ as described in the text.

\begin{figure}
\centering \includegraphics[width=0.5\textwidth]{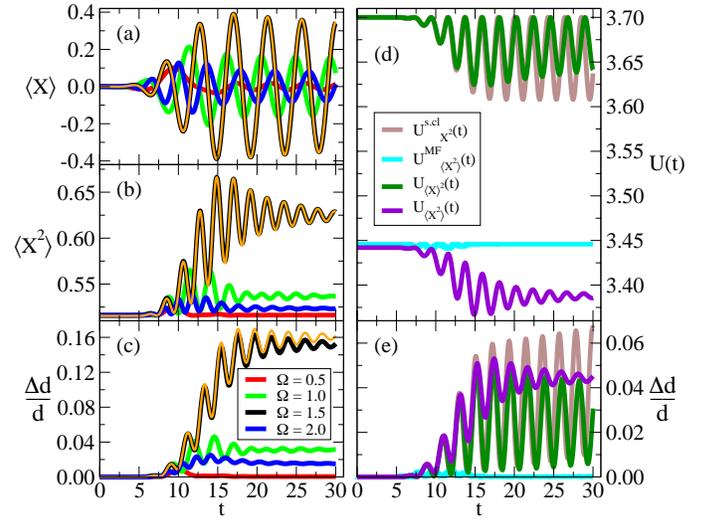}
\caption{ (Color online) Time evolution of the expectation value of the position operator $\langle X \rangle$ (a), its square $\langle X^{2} \rangle$ (b), the relative change of the double occupations with respect to the initial value $\frac{\Delta d}{d} = \frac{d \left( t \right) - d \left( 0 \right)}{d \left( 0 \right)}$ (c), at $U=3.7$, $\beta=10$, and field pulses $f(t)$ of different frequencies $\Omega$. The thin orange lines represent the results at $\Omega = 1.5$ without Ohmic bath. (d) Time dependent $U^{\text{s.cl}}_{X^{2}}(t)$, $U^{\text{MF}}_{\langle X^{2} \rangle}(t)$, $U_{\langle X\rangle^2}(t)$, and $U_{\langle X^2\rangle}(t)$, as defined in Eq.~\eqref{u_driv}. For the protocols $U_{\langle X\rangle^2}(t)$, and $U_{\langle X^2\rangle}(t)$ the expectation value $\langle X(t) \rangle$ and $\langle X^{2}(t) \rangle$ correspond to the black line at $\Omega = 1.5$ of panels a and b. (e) Time-dependent $\frac{\Delta d}{d}$ as obtained from \equ{ham_u_driv}, with $U \left( t \right)$ shown in panel d.}
\label{fig_2}
\end{figure}

\begin{figure*}
\centering \includegraphics[width=1\textwidth]{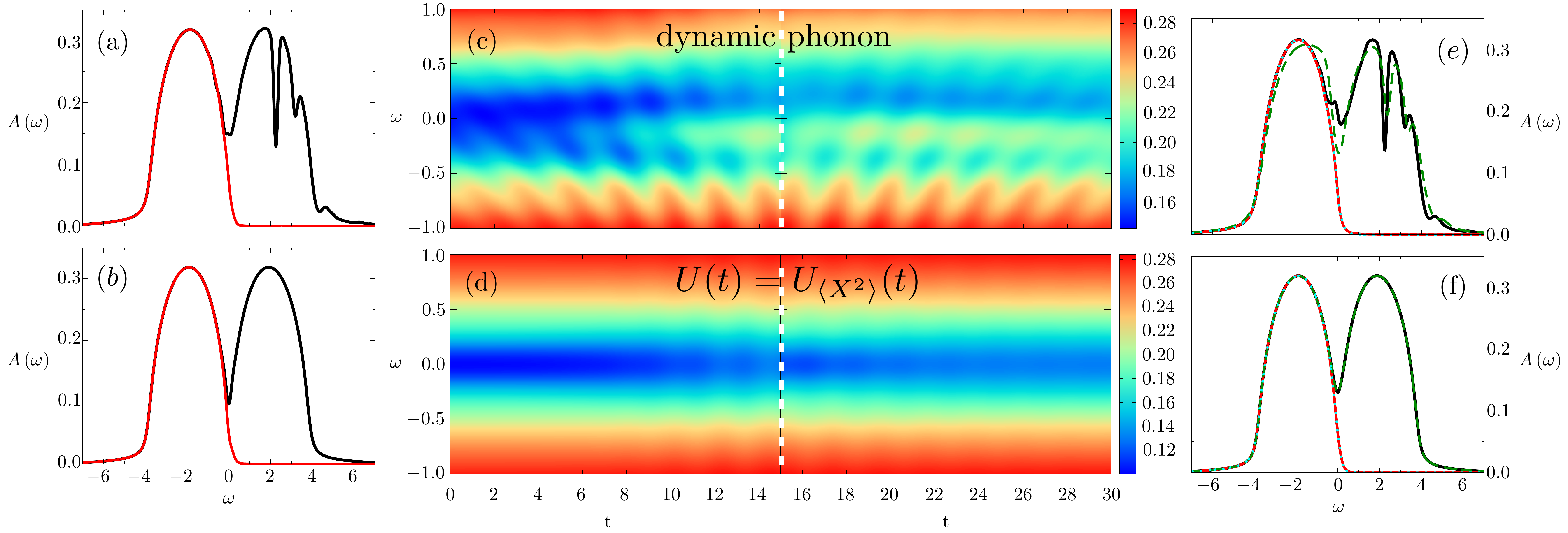}
\caption{ (Color online) Spectral functions for the dynamic phonon model at $U=3.7$ (upper row) and the simplified approach $U(t)=U_{\langle X^2\rangle}(t)$ (bottom row). Spectral function and occupation (black and red lines, respectively) at initial time $t=0$ and at latest time $t = 30$ are shown in the left (a-b) and right columns (e-f), respectively. For intermediate time-steps see \appu{appa}. Additional lines in (e) and (f) show a fit $A^<(\omega,t)=A(\omega,t)/(1+e^{\beta_{\rm eff}\omega})$ to the occupation functions (dotted cyan lines), and the equilibrium spectral function (dashed green lines) for a systems with the same total energy as the driven system at latest time $t=30$ ($\beta \sim 2.083$ for the phonon case, $U \sim 3.388$ and $\beta = 10$ for the $U$-driven case). Middle panels (c-d): Time dependent spectral weight close to the Fermi level $\omega = 0$; spectral functions are obtained by forward Fourier transform $A(\omega,t) =-\frac{1}{\pi}\text{Im}\int ds G^R(t+s,t) e^{i\omega s}$ for $t<15$ and by backward Fourier transform for $t>15$ (dashed vertical line at $t=15$).}
\label{fig_4}
\end{figure*}

We solve the time-dependent models \equ{ham} and \equ{ham_u_driv} using non-equilibrium 
dynamical mean-field theory (NEDMFT) within non-crossing approximation (NCA) \cite{Aoki2014,PhysRevB.82.115115}. 
NCA has limitations in the description of the metallic phase of the Hubbard model at low temperatures and small $U$. Here, we apply it at high temperature and intermediate values of $U$ ($U \sim$ bandwidth), where it is known to be qualitatively correct. 
NEDMFT maps the lattice problem into an Anderson impurity model with a self-consistent 
hybridization function $\Delta ( t,t' )$. We use a semi-elliptic free density of states of 
width $4v_0$, leading to the closed form $\Delta ( t,t' ) = v_{0}^{2} G ( t,t' )$ in terms of the 
local contour-ordered electronic Green's function $G( t,t' )$. For model \eqref{ham}, we 
include the full phonon Fock space and the nonlinear local electron-phonon dynamics 
exactly in the DMFT impurity model (similar to bosonic DMFT \cite{PhysRevX.5.011038}) 
and take the cutoff in the phonon Hilbert space large enough ($N_{\text{ph}} = 18$ for mean 
occupations $|\langle X \rangle| \lesssim 1$). This avoids potential ambiguities of alternative 
diagrammatic approaches for the local electron-phonon interaction \cite{PhysRevB.88.165108,doi:10.1063/1.4935245,PhysRevB.93.174309}. 
Finally, energy dissipation of electrons to other degrees of freedom (phonons, spin 
fluctuations, etc.), which is fast in correlated insulators, is taken into account through a 
bosonic heat bath. The bath just adds a term to the hybridization \cite{Eckstein2013,PhysRevB.101.161101}, 
$\Delta \left( t,t' \right) = v_{0}^{2} G \left( t,t' \right) + \Delta_{\text{Ohmic}} \left( t,t' \right)$, 
where $\Delta_{\text{Ohmic}} \left( t,t' \right) = \lambda G  \left( t,t' \right) D_{\text{Ohmic}}  \left( t,t' \right)$ 
is second-order in the electron-boson coupling with temperature $1/\beta$ and Ohmic 
density of states (bath spectral density $J \left( \omega \right) = \sum_{\alpha} g_{\alpha}^{2} \delta \left( \omega - \omega_{\alpha} \right) = \omega \theta \left(\omega_{\text{c}} - \omega \right)$, $\omega_{\text{c}} = 0.2$). We take $\lambda = 0.242$, which is weak enough so that 
electronic spectra are not affected by the coupling to the bath.


\textit{Results -} In \figu{fig_1}a, we plot the electronic spectral function $A \left( \omega \right)$ 
of the model \eqref{ham} for several values of the interaction $U$. 
At inverse temperature $\beta = 10$, shared by both the electronic and lattice subsystems, there is an IMT around $U = 3.6$, indicated by the appearance of a quasi-particle peak at $\omega \sim 0$. With increasing temperature, 
the quasi-particle peak vanishes, as the system crosses over into a bad metallic regime 
(see data for $U=3.0$ and $\beta=2$). The pronounced dip in the upper Hubbard band 
results from the hybridization between the doublon states and the phonon.

\begin{figure}
\centering \includegraphics[width=0.5\textwidth]{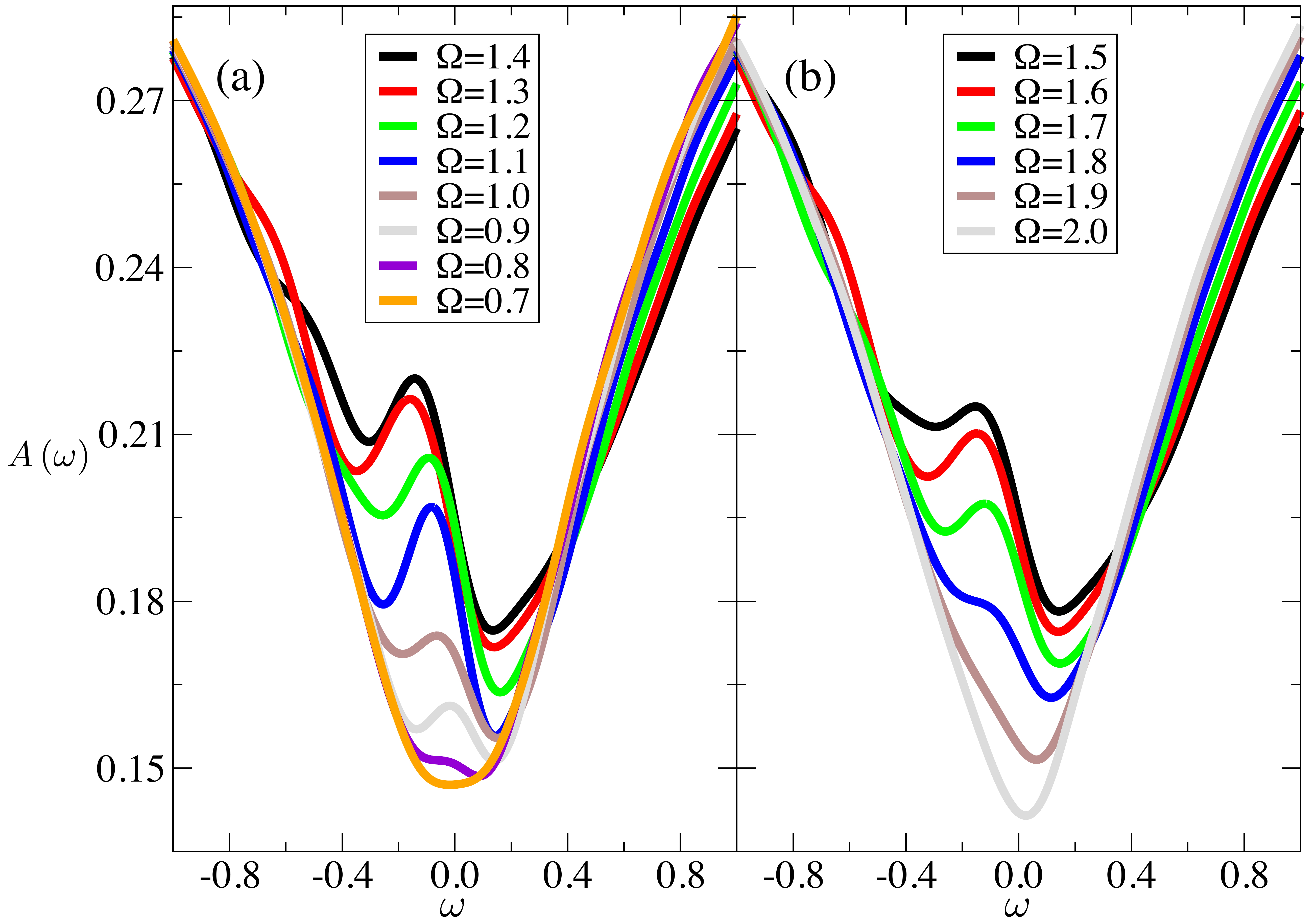}
\caption{ (Color online) Zoom around $\omega \sim 0$ of the time dependent (backward) spectral functions $A(\omega,t) =-\frac{1}{\pi}\text{Im}\int ds G^R(t,t-s) e^{i\omega s}$ at $t=30$ after excitation pulses with frequency $\Omega$ below (a) and above (b) the bare phonon frequency $\omega_{\text{ph}}$.}
\label{fig_3}
\end{figure}

We now concentrate on parameters $U = 3.7$ and $\beta = 10$ above the IMT 
and attempt to induce the transition through phonon driving. The vibrational mode 
is pumped using a few-cycle pulse $f(t)$ as shown in the inset of \figu{fig_1}, at 
different frequencies $\Omega$. Figure~\ref{fig_2} displays the resulting time-evolution 
of several observables. While $\langle X \rangle$ oscillates around its equilibrium 
position $\langle X \rangle = 0$ with the bare phonon frequency $\sim \omega_{\text{ph}}$ (\figu{fig_2}a), 
$\langle X^{2} \rangle$ oscillates around a shifted mean at a frequency $\sim 2 \omega_{\text{ph}}$ 
(\figu{fig_2}b). The most pronounced response is observed at resonant pumping 
$\Omega =  \omega_{\rm ph}=1.5$. At resonance, the double occupancy increases by 
about $15 \%$ compared to its initial value (Fig.~\ref{fig_2}c). These changes go along 
with a photo-induced metallicity, as observed from the transient appearance of a 
quasi-particle peak at $\omega \sim 0$ in the time-resolved spectral function 
(\figu{fig_4}, upper panels). The metallization survives the switch-off of the laser pulse 
up to the latest time ($t=30$) of the simulation (\figu{fig_4}e). In addition to the quasi-particle 
peak, one observes oscillations at frequency $2 \omega_{\text{ph}}$ in the spectrum, like
 in $\langle X^{2} \rangle$. The broad range of frequencies $\Omega$ where we observe 
 the onset of the peak at $\omega \sim 0$ verifies the genuine metallicity of the 
 photo-induced state, with a small asymmetry around the observed maximum response, 
 see Fig.~\ref{fig_3}.

We will now contrast these results with the decoupled model \equ{ham_u_driv}, 
where phonon variables in the electron-phonon interaction \equ{ham_el_ph} are 
replaced by a classical field or the expectation value of a quantum operator, 
leading to a time-dependent interaction. We compare four possible ansatzes: 
(i) $X_i^{2}$ is replaced by a classical field $X(t)^2$ self-consistently determined 
by the semiclassical equation of motion, (ii) a mean-field decoupling of the electron-phonon 
interaction \equ{ham_el_ph} (of the kind $O_{i} X^{2}_{i} \rightarrow \langle O_{i} \rangle X^{2}_{i} + \langle X^{2}_{i} \rangle O_{i} - \langle O_{i} \rangle \langle X^{2}_{i} \rangle$) 
is performed while the full quantum dynamics of phonons is considered with 
the equation of motion for the density matrix, and $X^{2}$ is substituted with 
(iii) $\langle X(t) \rangle^{2}$ or (iv) $\langle X^{2}(t) \rangle$ obtained from the 
full DMFT calculation. Those approaches lead to
\be \label{u_driv}
U \rightarrow 
\left\{
\begin{array}{c}
U - 2 \left( d-h \right) X^{2}(t) 
\equiv U^{\text{s.cl}}_{X^2}(t)
\\
U - 2 \left( d-h \right) \langle X^{2}(t) \rangle
\equiv U^{\text{MF}}_{\langle X^2\rangle}(t)
\\
U - 2 \left( d-h \right) \langle X(t) \rangle^{2}
\equiv U_{\langle X\rangle^2}(t)
\\
U - 2 \left( d-h \right) \langle X^{2}(t) \rangle
\equiv U_{\langle X^2\rangle}(t)
\end{array}
\right.,
\ee 
where we have assumed $\sum_i H_i =\sum_i D_i$ up to terms $\propto \hat N$. 
\figu{fig_2}d shows the resulting $U(t)$ for the almost-resonant driving case $\Omega=1.5$. 
The two self-consistent protocols are computed using the driving term shown in the inset of \figu{fig_1}.
Clearly, for a general state of the quantum phonon, $\langle X \rangle^{2} \neq \langle X^{2} \rangle$. 
A time-dependent function of the kind $U \left( t \right) = U + \Delta U \left[ 1 - \cos \left( 2 \omega_{\text{ph}} t \right) \right] \theta \left( t \right)$ 
can qualitatively describe the $U$-driving guided by the classical $X$ or by $\langle X \rangle$, while $U_{\langle X^2\rangle}(t)$ 
looks more like an interaction quench. 
Finally, $U^{\text{MF}}_{\langle X^2\rangle}(t)$, where the time-dependent $\langle X^{2}(t) \rangle$ is computed using the density matrix, shows a tiny time-dependence during the action of the external pulse only. Further details about the classical and mean-field dynamics are provided in \appu{appb}.

We note that $U^{\text{MF}}_{\langle X^2\rangle}$ and $U_{\langle X^2\rangle}$ deviate from $U$
already in equilibrium, where the 
vacuum and thermal fluctuations of the phonon are responsible for a renormalization 
of $U$ by $\sim -0.3$. In equilibrium, one finds that this renormalization rather 
accurately accounts for the shift of the IMT in the 
generalized Hubbard-Holstein model as 
compared to the standard Hubbard model. In Fig.~\ref{fig_1}b, we show the 
equilibrium spectra of the Hubbard model at different $U$; the IMT in the Hubbard-Holstein 
model \equ{ham} is indeed lowered by $\sim -0.3$ compared to the static 
one. This quantitative agreement also shows that for the given parameters polaronic 
effects play a minor role in localizing the quasi-particles, at least under equilibrium 
conditions. In contrast to this observation, the non-equilibrium 
dynamics of quasi-particles and the vibrationally induced IMT in the 
model cannot be explained by a renormalized time-dependent interaction. We compare 
in \figu{fig_4} the time-dependent spectra $A \left( \omega, t \right)$ for the 
full dynamical phonons simulation and the $U_{\langle X^2\rangle}$ driving protocol, that we take as representative of all the simplified protocols \equ{u_driv} (indeed, the qualitative dynamics of $A \left( \omega, t \right)$ is the same for all of them).
While we still notice some increase in the spectral weight at zero frequency, 
the quench-like $U(t)=U_{\langle X^2\rangle}(t)$ (lower panels) 
does not reproduce the emergence of a zero-frequency peak. The rather different response of 
full DMFT treatment and the simplified approaches \eqref{u_driv} 
is evident also in other observables: 
the relative change $\Delta d/d \sim 15\%$ in the double occupancy at resonant driving is 
almost three times larger than the change $\Delta d/d$ in the Hubbard model in response 
to the corresponding time-dependent interactions (cf.~Figs.~\ref{fig_2}c and e).

A possible explanation of the above findings is that the simplified approaches, which generally 
replace the phonon operators by semiclassical fields in the electronic problem, have underestimated the phonon-induced dissipation 
which favors metallization. However, we find that different electronic temperatures cannot explain our results. An effective temperature $1/\beta_{\text{eff}}$ obtained 
from a fit of the equilibrium fluctuation relation $A^<(\omega,t)=A(\omega,t)/(1+e^{\beta_{\rm eff}\omega})$ 
to the occupation functions $A^<(\omega,t)$ close to $\omega=0$ at the latest 
simulation time yields an even lower effective temperature for the $U_{\langle X^2\rangle}$-driven 
case ($\beta_{\text{eff}} \sim 9.5$) compared to the full DMFT results ($\beta_{\text{eff}} \sim 6.2$), 
see continuous red lines and dotted cyan lines in Figs.~\ref{fig_4}e and f. Moreover,  $1/\beta_{\text{eff}}$ 
obtained in this way can only characterize the low-energy quasi-particles, while 
the total energy at $t=30$ is that of a system at inverse temperature $\beta \sim 2.083$ (spectrum shown 
by the dashed green line in Fig.~\ref{fig_4}e), and $\langle X\rangle$ is still oscillating 
(Fig.~\ref{fig_2}a). We, therefore, conclude that a time-dependent interaction as in \equ{u_driv} 
plus electron cooling cannot faithfully describe the enhancement of metallicity 
in the photoexcited Mott insulator with nonlinear electron-phonon coupling.
An analogous analysis shows that even for states closer to the metal-insulator transition, the 
enhancement of metallicity is underestimated. A possible explanation is that 
in addition to the static renormalization of $U$ 
through $\langle X^2\rangle$, there is a dynamic contribution via virtual phonon emission 
and absorption. Such induced interactions go as $g^2/\omega_{\rm ph}$ in the anti-adiabatic phonon regime. In the 
presence of the $X^2$-nonlinearity, an oscillating phonon may act similar to a dynamically 
modulated electron-phonon coupling, which allows for interactions mediated via 
phonon-Floquet sidebands, with an energy denominator $1/(\omega_{\rm ph}\pm\Omega)$. 
Thus, such interactions can be strongly renormalized, in particular close to resonance 
$\omega_{\rm ph}=\Omega$. While the present parameter regime, close to the IMT and 
with not too well-separated energy scales between $\omega_{ph}$ and bandwidth, 
makes an analytic understanding difficult, the numerical results in Fig.~\ref{fig_3} 
unambiguously demonstrate substantial enhancement of the quasi-particle peak around 
the resonance. Another effect could be driving-induced undressing of polarons \cite{PhysRevB.66.064507,PhysRevB.47.5351,Gaal2007}, 
but as we have concluded above, polaronic effects are probably not significant here.


\textit{Conclusions -} In this study, we have analyzed a generalized 
Hubbard-Holstein model, relevant for the description of molecular 
solids, which is excited by an ultrafast pulse of local molecular vibration. 
By exactly treating the quantum phononic fluctuations, we show numerical evidence 
for the vibrationally-induced emergence of a quasi-particle peak in the electronic 
spectral function at $\omega \sim 0$, signalling the occurrence of an IMT. This 
observation should be relevant for the understanding of photo-induced 
superconductivity in molecular solids \cite{buzzi2020photomolecular}. More generally, 
the striking difference of our results to a simplified treatment for the phonons 
implies the need for careful analysis of quantum phonon effects in the phononic 
control of electronic properties.

F.G. would like to thank Nagamalleswararao Dasari for the useful discussions about the 
Ohmic bath. We were supported by the ERC starting grant No. 716648. The authors 
gratefully acknowledge the computational resources and support provided by the 
Erlangen Regional Computing Center (RRZE).


\appendix

\section{Spectral functions and occupations at intermediate time-steps} \label{appa}

\begin{figure*}
\centering \includegraphics[width=1\textwidth]{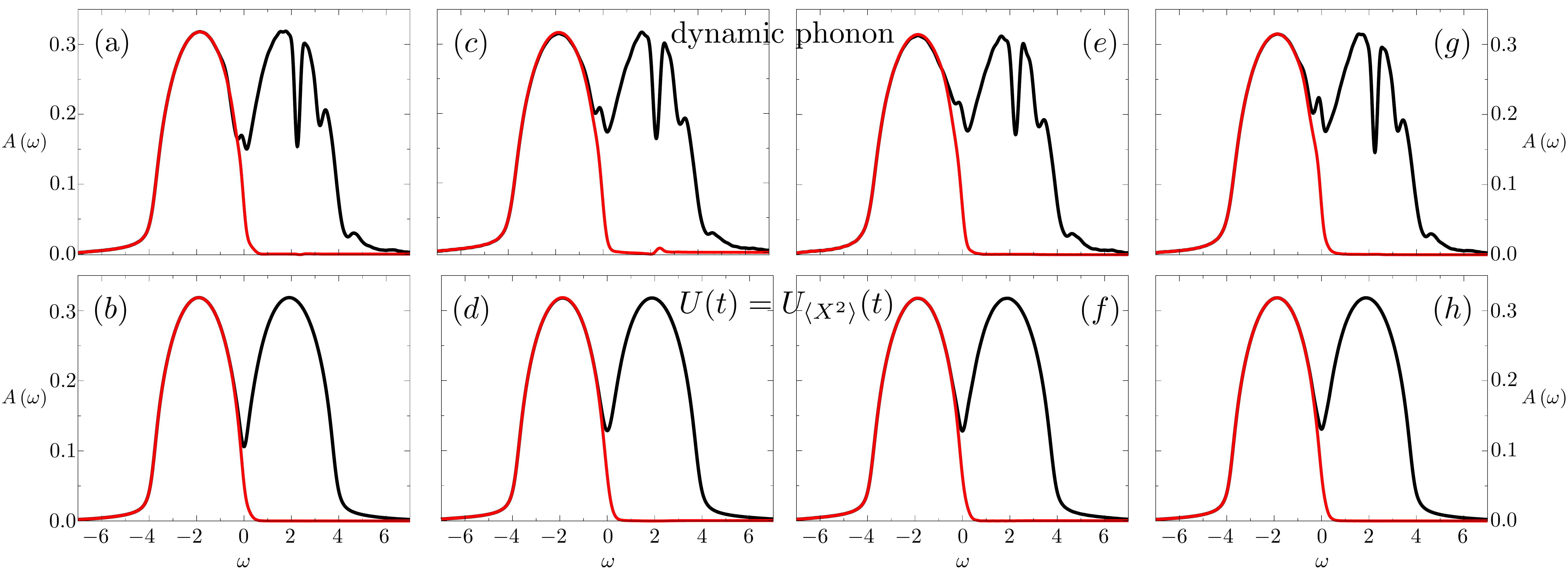}
\caption{ (Color online) Time-dependent spectral functions for the dynamic phonon model at $U=3.7$ (upper row) and for the simplified model with $U(t)=U_{\langle X^{2} \rangle}(t)$ (bottom row) as in \figu{fig_4} of the main text. Spectral function and occupation (black and red lines, respectively) at time $t=6$ (panels (a-b)), $t = 12$ (panels (c-d)), $t=18$ (panels (e-f)) and $t=24$ (panels (g-h)).}
\label{fig_suppl_mat_fig_1}
\end{figure*}

In \figu{fig_4} of the main text, we have shown the time-dependent 
spectral functions at the initial ($t=0$) and final ($t=30$) times 
of the simulations we performed for the dynamic phonon model 
and the simplified approach based on the time-dependent Hubbard interaction $U_{\langle X^2\rangle}(t)$. 
In the same figure, we have also presented the time evolution of 
the spectral weight around the Fermi level. Here, we want to 
show the spectral functions and the respective occupations at 
intermediate times for the two 
different cases as depicted in 
\figu{fig_suppl_mat_fig_1}. We notice that the most significant 
changes of the spectral functions occur around $\omega \sim 0$ 
and that we observe an overall increase of the spectral weight 
at the Fermi level  
in both cases (and, actually, also for the other simplified treatments of the electron-phonon interaction presented in \equ{u_driv} of the main text). 
For the dynamic 
phonon, we recognize the presence of a quasi-particle peak 
already at $t=6$. The dynamics of this peak looks pretty 
interesting: by comparing the snapshots of the spectral 
function taken at different times, we distinguish the breathing 
of the peak in both its height and width. The 
$U_{\langle X^2\rangle}(t)$ driving leads instead to an almost featureless 
dynamics, with just a 
small and featureless increase of the weight at $\omega = 0$ 
and a slight broadening of the occupation function. 


\section{Semiclassical approximation and mean-field decoupling of the electron-phonon interaction} \label{appb}

In the main text, we have presented four
alternative ways of 
replacing, in the electron-phonon interaction, the phonon 
degree of freedom with a classical field 
or with a simplified treatment of the quantum phonons, see \equ{u_driv} appearing 
there. 
While the two driving protocols of the purely electronic model $U_{\langle X\rangle^{2}}(t)$ and $U_{\langle X^2\rangle}(t)$ do not involve in any sense the phonon part of the Hamiltonian, since the expectation values $\langle X\rangle$ and $\langle X^2\rangle$ are obtained from the full DMFT calculation, the semiclassical and the mean-field approaches still retain the phonon degree of freedom. In this sense, these last two treatments take into account the back-action of the electrons on the displacement field. 
The first approach we describe is the semiclassical one, based on 
the replacement of the quantum operator $X_{i}$ with a 
classical and site-independent field $X$. 
Similarly, we introduce the classical field $P$, the conjugate momenta of the generalized coordinate $X$.

To compute their time evolution, we use the Hamilton equations of motion:
\be \label{class_syst}
\left\{
\begin{array}{c}
\dot{X} \left( t \right) = \omega_{\text{ph}} P \left( t \right) \ \ \ \ \ \ \ \ \ \ \ \ \ \ \ \ \ \ 
\\
\dot{P} \left( t \right) = - \frac{\omega_{\text{ph}}^{2} \left( d \left( t \right) \right)}{\omega_{\text{ph}}} X \left( t \right) - \frac{F \left( t \right)}{\omega_{\text{ph}}}
\end{array}
\right.,
\ee
where $\omega_{\text{ph}}^{2} \left( d \left( t \right) \right) = \omega_{\text{ph}}^{2} \left[ 1 + \frac{4 \left( h - d \right)}{  \omega_{\text{ph}} } d \left( t \right) \right]$, 
with $d \left( t \right) = \frac{1}{N} \sum_{i} \langle D_{i} \rangle$ 
being the time-dependent expectation value of the double 
occupations. $F \left( t \right) = \sqrt{2} \omega_{\text{ph}}^{2} f \left( t \right)$ 
is a force field related to external driving. The solution 
of the system \equ{class_syst} provides the 
time-dependence of the fields $X \left( t \right)$ and 
$P \left( t \right)$ so that we can write the time-dependent 
electronic model that we have to deal with as:

\be \label{ham_el_time_dep}
H_{\text{U-driv}} \left( t \right) = H_{\text{el}} - 2 X^{2} \left( t \right) \left( d-h \right) \sum_{i} D_{i} \;,
\ee
where $H_{\text{el}}$ is the electronic part of the original Hamiltonian defined in \equ{ham_el} of the main text.

This way, we obtain a time-dependent Hubbard 
interaction written as:

\be \label{u_classical}
U_{X^{2}}^{\text{s.cl}} \left( t \right) = U - 2 \left( d-h \right) X^{2} \left( t \right) \;.
\ee

The Hamiltonian \equ{ham_el_time_dep} corresponds to \equ{ham_u_driv} in the main text, where the time-dependent Hubbard interaction is provided by \equ{u_classical}. This equation has to be supplied with \equ{class_syst}, that provides the time dependence of the field $X \left( t \right)$.
The dependence of \equ{class_syst} by the time-dependent 
double occupations $d \left( t \right)$ leads to a back-action 
of the electrons on the classical phonon. 

\begin{figure}
\centering \includegraphics[width=0.5\textwidth]{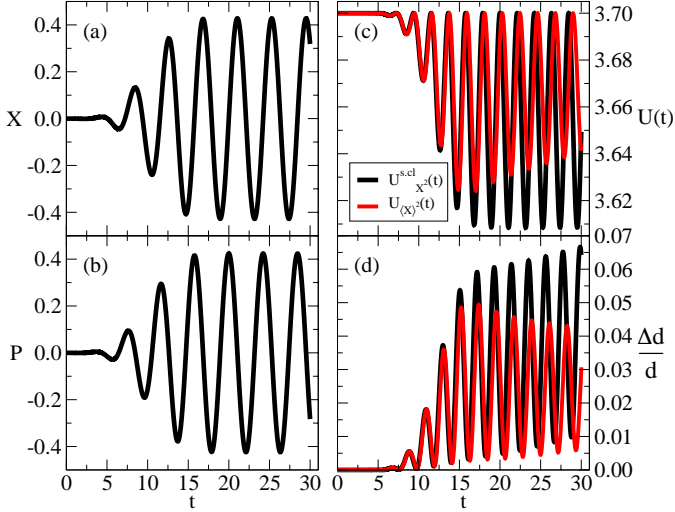}
\caption{ Time evolution of the classical fields $X \left( t \right)$ and $P \left( t \right)$ (panels (a) and (b), respectively) and the corresponding changes in the Hubbard interaction $U_{X^{2}}^{\text{s.cl}} \left( t \right)$ and in the relative change in the double occupations $\Delta d / d$ (panels (c) and (d), respectively) for the same excitation protocol shown in the inset of \figu{fig_1} of the main text (black lines). For comparison, in panels (c) and (d) we show the results for the time dependent protocol $U_{\langle X \rangle^{2}} \left( t \right)$ (red lines). The lines in panels (c) and (d) are already shown in \figu{fig_2}(d) and (e) of the main text.}
\label{fig_suppl_mat_fig_2}
\end{figure}

By driving the classical field $X \left( t \right)$ with an excitation protocol $f \left( t \right)$ as depicted in the inset of Fig.1 of the main text, and by using the same Hamiltonian parameters defined there, we obtain the results shown in \figu{fig_suppl_mat_fig_2} (black lines). For comparison, we also show the results obtained for the $U_{\langle X \rangle^{2}} \left( t \right)$ driving already discussed in the main text (red lines).

\figu{fig_suppl_mat_fig_2}(a) ((b)) shows the time evolution of $X \left( t \right)$ ($P \left( t \right)$). The two fields oscillate out of phase at the same frequency close to the bare phonon one $\omega_{\text{ph}}$, and they keep doing it even when the pulse is over (we remind that the pulse duration is $20$). \figu{fig_suppl_mat_fig_2}(c) compares the $U$-driving experienced by the electronic part of the system during the semiclassical dynamics $U^{\text{s.cl}}_{X^{2}} \left( t \right)$, and for the $U_{\langle X \rangle^{2}} \left( t \right)$ driving we get from $\langle X \left( t \right) \rangle$ obtained with the full DMFT calculation. The comparison between the two gives a qualitative similarity of the results.

Not surprisingly, within the semiclassical phonon treatment, we cannot get two independent renormalizations of the Hubbard interaction as we might get, at the quantum level, from $\langle X \rangle^{2}$ and $\langle X^{2} \rangle$. The reason for this is that, trivially, at the classical level we can simply compute $X^{2} \left( t \right)$ by taking the square of $X \left( t \right)$, while this procedure dramatically fails at the quantum level. Thus, to describe a driving protocol based on a field $X^{2}$ with an expectation value independent from the one of $X$, we must rely on a quantum-mechanical description of the operator $X^{2}$. A possible way to keep the quantum nature of the bosonic field is to perform a mean-field decoupling of the electron-phonon interaction presented in \equ{ham_el_ph} of the main text. This reads:

\be \label{mean_field}
\begin{split}
H_{\text{el-ph}} \rightarrow & 2 \left( h - d \right) d \left( t \right) \sum_{i} X^{2}_{i} \\
& + 2 \left( h - d \right) \langle X^{2} \rangle \sum_{i} D_{i} \\
& - 2 N \left( h - d \right) \langle X^{2} \rangle d \left( t \right) \;.
\end{split}
\ee

This way, we can separately write the electronic and 
phononic mean-field Hamiltonians, coupled one to the 
other, as:

\be \label{mean_field_ham}
\begin{split}
& H_{\text{el}}^{\text{MF}} = H_{\text{el}} + 2 \left( h - d \right) \langle X^{2} \left( t \right) \rangle \sum_{i} D_{i} \;, \\
& H_{\text{ph}}^{\text{MF}} = H_{\text{ph}} + H_{\text{driv}} \left( t \right) + 2 \left( h - d \right) d \left( t \right) \sum_{i} X^{2}_{i} \;,
\end{split}
\ee

where we omitted the term $- 2 N \left( h - d \right) \langle X^{2} \rangle d \left( t \right)$ 
of \equ{mean_field}.

We solve the electronic part of the 
model $H_{\text{el}}^{\text{MF}}$ with DMFT at the 
NCA level. Since the phononic part of the mean-field Hamiltonian is local, we are allowed to write $H_{\text{ph}}^{\text{MF}} = \sum_{i} H_{\text{ph},i}^{\text{MF}}$. The time evolution of the phonon degree of freedom might thus be computed using the density matrix, that at equilibrium, locally, looks:

\be
\rho_{\text{ph},i}^{\text{eq}} = \frac{e^{- \beta H_{\text{ph},i}^{\text{MF}}}}{\text{Tr} \left[ e^{- \beta H_{\text{ph},i}^{\text{MF}}} \right]} = V \frac{e^{- \beta H_{\text{ph},i, \text{d}}^{\text{MF}}}}{\text{Tr} \left[ e^{- \beta H_{\text{ph},i, \text{d}}^{\text{MF}}} \right]} V^{\dagger} \;,
\ee

where $H_{\text{ph},i}^{\text{MF}} = V_{i} H_{\text{ph},i,d}^{\text{MF}} V_{i}^{\dagger}$ and $H_{\text{ph},i,d}^{\text{MF}}$ 
is the diagonal form of the local mean-field phononic 
Hamiltonian $H_{\text{ph},i}^{\text{MF}}$. We 
underline that, at equilibrium, $H_{\text{driv}} \left( t \right)$ 
is equal to zero. The subsequent time-evolution of 
the density matrix can be computed via the 
Von Neumann equation:

\be
\frac{\partial \rho_{\text{ph},i} \left( t \right)}{\partial t} = -i \left[ H_{\text{ph},i}^{\text{MF}}, \rho_{\text{ph},i} \left( t \right) \right] \;,
\ee

with initial condition provided by $\rho_{\text{ph},i} \left( t=0 \right) = \rho_{\text{ph},i}^{\text{eq}}$.

A convenient basis for expressing the density matrix, as 
well as $H_{\text{ph}}^{\text{MF}}$, is the one of the 
local phonon Fock space so that we can write:

\be
\begin{split}
& \left( H_{\text{ph},i} \right)_{n,p} = \omega_{\text{ph}} N \left( p + \frac{1}{2} \right) \delta_{n,p} \;, \\
& \left( H_{\text{driv},i} \left( t \right) \right)_{n,p} = \omega_{\text{ph}} N f \left( t \right) \left[ \sqrt{p+1} \delta_{n, p+1} + \sqrt{p} \delta_{n, p-1} \right] \;, \\
& 2 N \left( h - d \right) d \left( t \right) \left( X^{2}_{i} \right)_{n,p} = N \left( h - d \right) d \left( t \right) \\
& [ \sqrt{p \left( p-1 \right)} \delta_{n, p-2} + \left( 1+ 2p \right) \delta_{n,p} \\
& + \sqrt{\left( p+1 \right) \left( p+2 \right)} \delta_{n, p+2} ] \;.
\end{split}
\ee

\begin{figure}
\centering \includegraphics[width=0.5\textwidth]{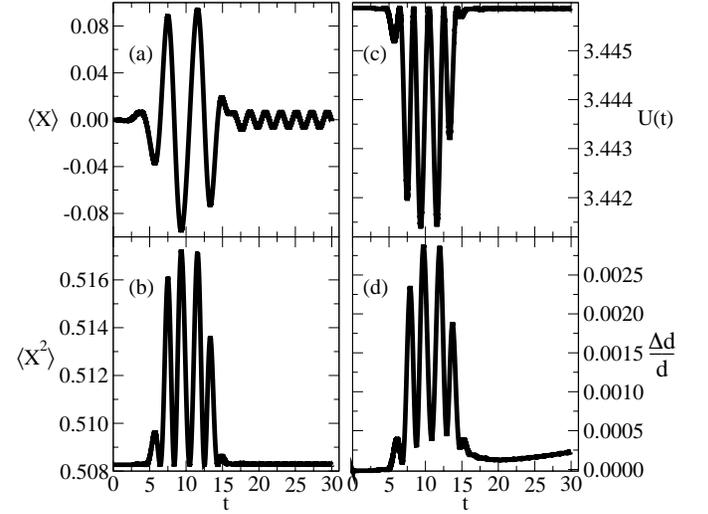}
\caption{ Time evolution of the expectation values $\langle X \left( t \right) \rangle$ and $\langle X^{2} \left( t \right) \rangle$ (panels (a) and (b), respectively) and the corresponding changes in the Hubbard interaction $U_{\langle X^{2} \rangle}^{\text{MF}} \left( t \right)$ and in the relative change in the double occupations $\Delta d / d$ (panels (c) and (d), respectively) for the same excitation protocol shown in the inset of \figu{fig_1} of the main text. We stress that the curves in panels (c) and (d) are shown also in \figu{fig_2}(d) and (e) of the main text, respectively.}
\label{fig_suppl_mat_fig_3}
\end{figure}

From the knowledge of the time-dependent density 
matrix $\rho_{\text{ph},i} \left( t \right)$, we compute 
the time-dependent expectation value of $X^{2}_{i}$ as:

\be
\langle X^{2} \left( t \right) \rangle  = \text{Tr} \left[ \rho_{\text{ph},i} \left( t \right) X^{2}_{i} \right] \;.
\ee

Given this quantity, we can also compute:
\be \label{u_mean_field}
U_{\langle X^{2} \rangle}^{\text{MF}} \left( t \right) = U - 2 \left( d-h \right) \langle X^{2} \left( t \right) \rangle \;,
\ee
defined in \equ{u_driv} of the main text.

In \figu{fig_suppl_mat_fig_3}, we show the results 
obtained at the mean-field level by considering 
the quantum phonons in the presence of an external driving equal to the one shown in the inset of \figu{fig_1} of the main text.
We notice that $\langle X^{2} \left( t \right) \rangle = \langle X \left( t \right) \rangle^{2}  + \langle X^{2} \left( 0 \right) \rangle$, 
where $\langle X^{2} \left( 0 \right) \rangle \sim 0.5082$. 
The time dependence of $\langle X \left( t \right) \rangle$ 
and of $\langle X^{2} \left( t \right) \rangle$, shown in 
\figu{fig_suppl_mat_fig_3}(a) and (b), respectively, does 
not resemble the one presented in the main text in 
panels (a) and (b) of \figu{fig_2} (black lines) 
and neither the one shown in \figu{fig_suppl_mat_fig_2}(a) for the classical field $X$. 
Indeed, in this 
mean-field calculation, we observe that the oscillatory 
behavior of both $\langle X \left( t \right) \rangle$ and 
$\langle X^{2} \left( t \right) \rangle$ is the most 
pronounced while the external pulse is active. Instead, when the external perturbation is over, 
the response of the system is strongly suppressed. 
The frequency of the oscillations observed in $\langle X \left( t \right) \rangle$ for $t > 20$ is almost equal to the double of the bare phonon frequency $2 \times \omega_{\text{ph}}$. This fact, together with the small response of the system for a driving frequency $\Omega = \omega_{\text{ph}}$, are in qualitative agreement with the picture provided by the parametric oscillator.
The quench in $\langle X^{2} \left( t \right) \rangle$ after the pulse 
observed in the full NCA calculation here disappears. 
Also, the relative change in the double occupations in 
\figu{fig_suppl_mat_fig_3}(d) is different as 
compared to the one presented in \figu{fig_2}(c) of the 
manuscript (black line) both from the quantitative and 
the qualitative point of view.

To briefly summarize our findings, we observe that 
the simplified protocol $U_{\langle X^{2} \rangle} \left( t \right)$ 
introduced in the main text (violet line in \figu{fig_2}(d)) 
produces a $\Delta d/d$ that compares much better 
to the NCA result with respect to the ones obtained 
with all the other simplified $U$-drivings introduced in \equ{u_driv} of the main text.



%

\end{document}